\documentclass[reprint,amssymb,superscriptaddress,nofootinbib]{revtex4-1}

\usepackage{xcolor}

\definecolor{dgreen}{RGB}{0,140,0}

\usepackage{tcolorbox}
\usepackage{amsfonts}
\usepackage{dsfont}
\usepackage{bm}
\usepackage{times}
\usepackage{verbatim}
\usepackage[utf8]{inputenc}
\usepackage[T1]{fontenc}
\usepackage[normalem]{ulem}
\usepackage[bottom]{footmisc}
\usepackage{amsmath}
\usepackage{amssymb}
\usepackage{mathrsfs}
\usepackage{mathtools}
\usepackage{amsthm}
\usepackage{physics}
\usepackage{bbm}

\usepackage{csquotes}

\makeatletter
\def\maketag@@@#1{\hbox{\m@th\normalfont\normalsize#1}} 
\makeatother

\usepackage{graphicx}
\usepackage{float}
\usepackage{subfigure} 
\usepackage{dcolumn}

\usepackage{pgfplots}
\usepgfplotslibrary{groupplots}

\DeclarePairedDelimiterXPP{\sfTr}[1]{\mathsf{Tr}}{[}{]}{}{#1}
\DeclarePairedDelimiterXPP{\sfTrAbs}[1]{\mathsf{TrAbs}}{[}{]}{}{#1}
\DeclarePairedDelimiterXPP{\opTr}[1]{\mathrm{Tr}}{[}{]}{}{#1}
\DeclarePairedDelimiterXPP{\bbTr}[1]{\mathbb{T}\mathrm{r}}{[}{]}{}{#1}
\DeclarePairedDelimiterXPP{\opTrAbs}[1]{\mathrm{TrAbs}}{[}{]}{}{#1}

\def\ANU{Centre for Quantum Computation and Communication Technology, Department of Quantum Science, Australian National University, Canberra, ACT 2601, Australia.}

\def\astar{Institute of Materials Research and Engineering, Agency for Science Technology and Research (A*STAR), 2 Fusionopolis Way, 08-03 Innovis 138634, Singapore}

\begin{document}

%\title{On the finite-copy attainability of the Holevo Cram{\'{e}}r-Rao bound and quantum Fisher information}
%\title{On the finite-copy attainability of the ultimate precision limits in quantum multiparameter estimation}
\title{Strong cubic phase shifts on the photonic vacuum state}
%\author{LC, JS, PKL, SMA}
\author{Hao Jeng}
\affiliation{\ANU}
\author{Lorc\'{a}n O. Conlon}
%\email{lorcanconlon@gmail.com}
\affiliation{\ANU}
\affiliation{\astar}
\author{Ping Koy Lam}
\affiliation{\ANU}
\affiliation{\astar}
\author{Syed M. Assad}
%\email{cqtsma@gmail.com}
\affiliation{\ANU}
\affiliation{\astar}
%\affiliation{\NTU}

\begin{abstract}
Addition of photons to coherent states is shown to produce effects that display remarkable similarities with cubic phase shifts acting on the vacuum state, with fidelities in excess of 90 percent. The strength of the cubic interaction is found to vary inversely with the displacement of the coherent state and the strongest interactions were one order of magnitude greater than previous observations. The interaction is non-perturbative.
\end{abstract}
\maketitle

\begin{figure*}[t]
\includegraphics[width=\textwidth]{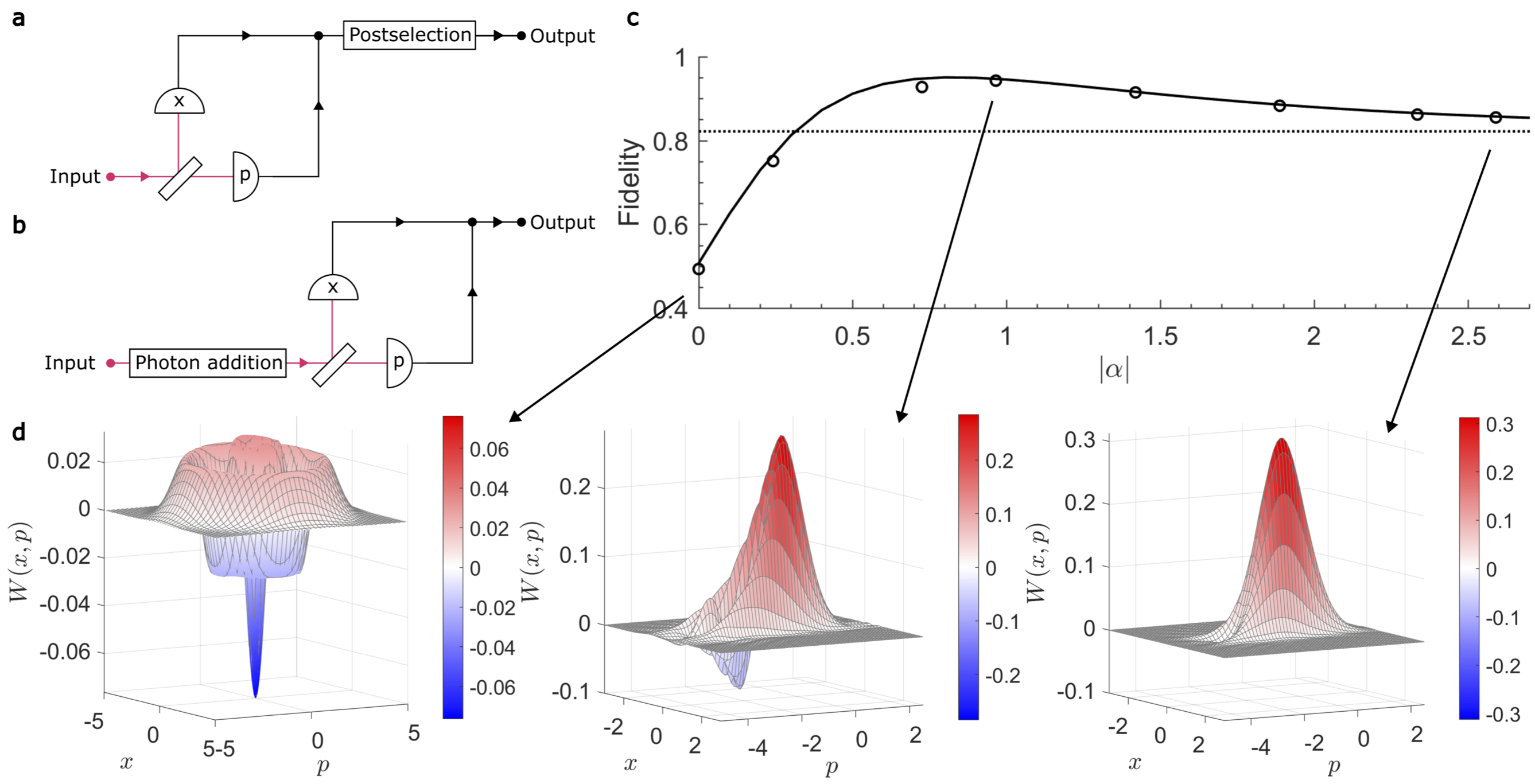}
\caption{\textbf{a.} Simplified schematic of the experiment. The input field is a single frequency laser at 1064 nm, carrying amplitude and phase modulations at the 4 MHz sideband as coherent states. The beamsplitter splits the beam equally, and the labels $x$ and $p$ indicate homodyne measurements of the amplitude and phase quadratures respectively. The probability of selecting each measurement is proportional to $(x^2 + p^2)^k$ in the region $x^2 + p^2 \leq 6$, with $k$ the number of photons added, and fixed to unity outside that region. \textbf{b.} An alternative but entirely equivalent representation of the experiment with postselection replaced by physical photon addition. \textbf{c.} Experimentally recorded fidelities between a fixed cubic state with interaction strength $\gamma = 0.4$ and coherent states with three added photons as a function of the complex amplitude of the coherent state. Dotted line indicates the maximum fidelity between Gaussian states and the $\gamma = 0.4$ cubic state. Each data point is obtained from about $10^8$ measurements of each quadrature. The probability of success in these experiments are approximately 0.001, depending principally on the number of photons added which has been fixed to three. \textbf{d.} Experimentally reconstructed Wigner functions for selected observation runs. The Wigner functions are obtained from density matrices truncated on the fortieth photon component, which in turn is obtained through maximum likelihood estimation. }
\label{fig1}
\end{figure*} 

Unitary operations acting on quantum states of light take the form $e^{\mathrm{i}\hat{H}s}$, where $\hat{H}$ is the Hamiltonian associated with the unitary and $s$ is a number describing the interaction strength~\cite{ref1}. In quantum optics, this Hamiltonian is typically a polynomial function of the photon creation and annihilation operators, $\hat{a}$ and $\hat{a}^\dagger$. The simplest interactions of this type are the Gaussian operations with linear and quadratic Hamiltonians, such as displacement and squeezing. These sort of interactions are familiar to quantum optics by now and can be implemented straightforwardly, requiring, at most, the second order optical nonlinearity.

The cubic phase shift is a unitary operation with a nonlinear Hamiltonian of the lowest possible order, defined as
\begin{equation}
\hat{U} = e^{\mathrm{i}\gamma \hat{x}^3}\;,
\label{eq1}
\end{equation}
where the operator $\hat{x}= \hat{a}+\hat{a}^\dagger$ denotes the amplitude quadrature and $\gamma$ the cubic interaction strength. The Hamiltonian is cubic and thus requires optical nonlinearities of higher order. Unlike Gaussian operations, the cubic phase shift creates non-classical states with Wigner functions distorted by bends and possessing negativities, so that, consequently, it is significantly more difficult to realize.

The cubic phase shift is one of the most fundamental unitary operations in quantum optics, and its importance is highlighted by its applications in the field of quantum computation. The set of gates comprising of single mode Gaussian operations, the cubic phase shift, and a two-mode interaction forms a universal gate set for quantum computation with continuous variables~\cite{ref2} --in fact, the cubic interaction (or rather, the nonlinear interaction) is also necessary, because if it is dropped then those that remain can be simulated efficiently with a classical computer~\cite{ref3}. In other words, the cubic phase shift is an essential ingredient for a particular route towards universal quantum computation.

There have been a number of theoretical suggestions for implementing the cubic phase shift. However, all of these approaches make at least one of the two following assumptions: (1) that the interaction is very weak
$(\gamma\ll1)$, so that it can be implemented in a perturbative fashion~\cite{ref4,ref5,ref6},
\begin{equation}
\hat{U}\approx 1 + \mathrm{i}\gamma\hat{x}^3\;,
\end{equation}
or (2) that advanced techniques such as photon counting~\cite{ref7,ref8} or the Kerr interaction~\cite{ref9} are readily available.

The pioneering experiments of Yukawa \textit{et al.}~\cite{ref10} have
shown that a projection $a_0\bra{0} + a_1 \bra{1} + a_2 \bra{2} + a_3 \bra{3}$
by measurements on one beam of a two-mode-squeezed-
vacuum state $\sum_{n=0}^{\infty} \lambda^n \ket{n}\ket{n}$ collapses the other beam into a controllable superposition of up to three photons. The resulting quantum state can be made to match that of the weak cubic state,
\begin{equation}
(1 + \mathrm{i}\gamma\hat{x}^3) \ket{0} = \ket{0} + 3\mathrm{i}\gamma \ket{1} + \sqrt{6}\mathrm{i}\gamma\ket{3}\;,
\label{eq3}
\end{equation}
by adjusting the coefficients $a_0$, $a_1$, $a_2$, and $a_3$ in the measurement.

The primary limitation of Yukawa’s method is its perturbative nature. In order to produce a quantum state with non-zero amplitudes in the higher photon numbers, a substantially more complicated measurement device is required. On the other hand, there is no upper limit on the number of photons produced by the general, non-perturbative cubic phase shift \mbox{Eq. ~\eqref{eq1}} as $\gamma$ goes to infinity.

In this letter, we describe a method for producing non-perturbative cubic states using the technique of photon addition, implemented via a probabilistic process. In brief, the experimental arrangement we used is the following. Coherent states $\ket{\alpha}$ are produced on a continuous-wave laser by amplitude and phase modulations at a fixed frequency, and a heterodyne measurement $\bra{\beta}$ is made on this state. The outcomes of the heterodyne measurement are postselected with acceptance probabilities proportional to $\abs{\beta}^{2k}$ to implement the addition of $k$ photons~\cite{ref11}, viz.,
\begin{equation}
\bra{\beta} (\hat{a}^\dagger)^k\hat{\rho}\hat{a}^k\ket{\beta} = \abs{\beta}^{2k} \bra{\beta} \hat{\rho}\ket{\beta}\;.
\end{equation}
The overall process is therefore represented by the expression:
\begin{equation}
\bra{\beta} (\hat{a}^\dagger)^k \ket{\alpha}\;.
\end{equation}
The heterodyne measurements provide sufficient information for tomographic reconstruction of the state using the maximum likelihood method of Lvovsky~\cite{ref12}.

Figure~\ref{fig1} shows experimental results with three photons added to coherent states of varying complex amplitudes. The reason for choosing to add three photons will become clear after reviewing the experimental results. The fidelity is evaluated with respect to a cubic interaction of fixed interaction strength,
\begin{equation}
F = \mel{\psi_\text{cubic}}{\hat{\rho}}{\psi_\text{cubic}}\;,
\end{equation}
where the reference cubic state is given by the cubic unitary acting on the vacuum state and defined only up to arbitrary Gaussian operations --in other words, a displaced squeezed cubic state.

\begin{figure*}[t]
\includegraphics[width=\textwidth]{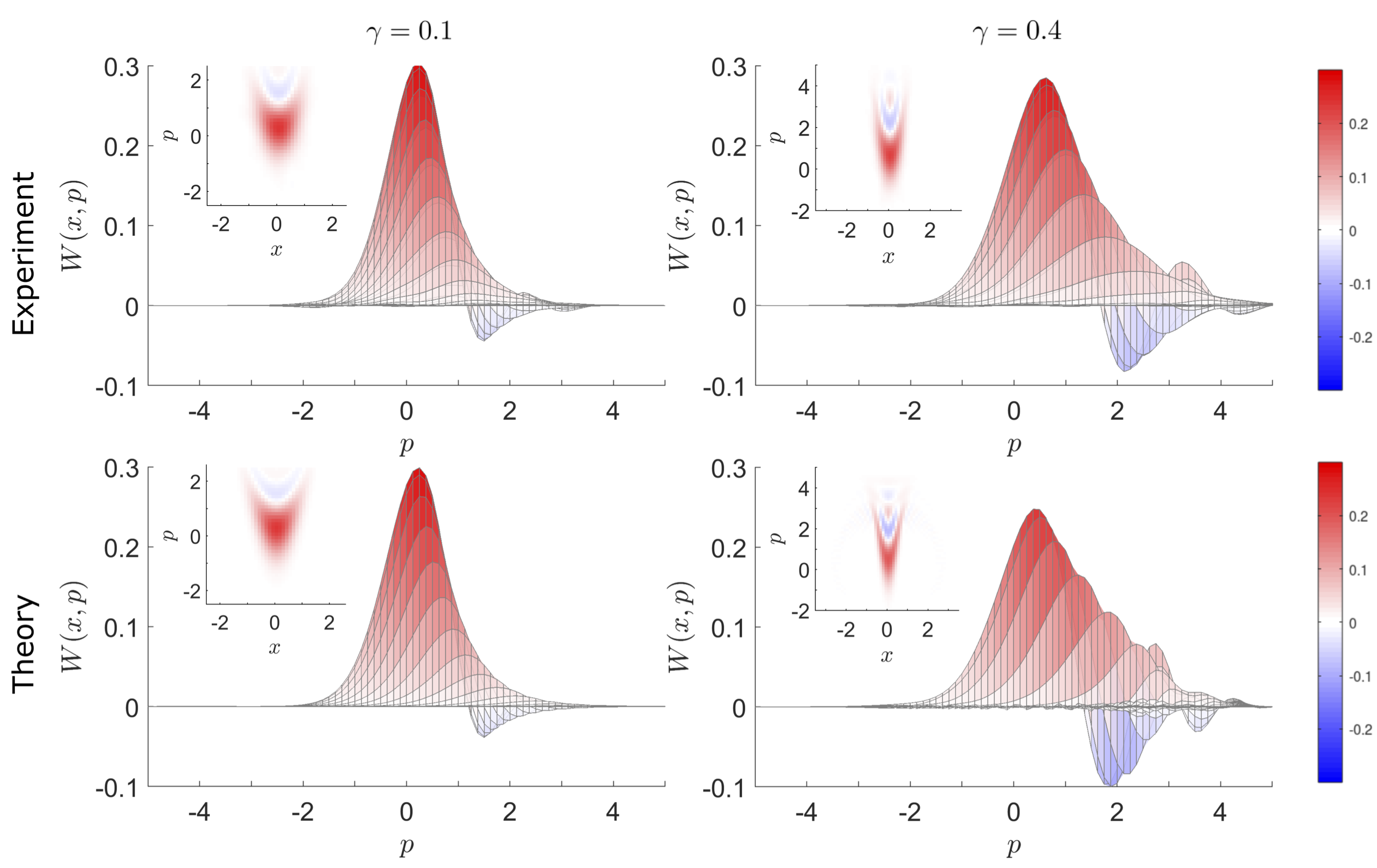}
\caption{Comparison of experimental and theoretical Wigner functions for interaction strengths of $\gamma = 0.1$ (left) and $\gamma = 0.4$ (right). Insets give top view of Wigner function. Complex amplitudes of the coherent states are $-1.47\mathrm{i}$ (left) and $-0.97\mathrm{i}$ (right) respectively for the two cases. All figures share the same colourbar.}
\label{fig2}
\end{figure*}

We observed a fidelity of 0.943 relative to a cubic interaction strength of $\gamma = 0.4$ at $\abs{\alpha} = 0.97$ (Fig.~\ref{fig1}), with a close-up of the corresponding Wigner function provided in Fig.~\ref{fig2}. For comparison, the fidelity relative to Gaussian states (optimised across displacement and squeezing by a grid search) is given by the horizontal line in Fig.~\ref{fig1} which is about 82 percent. We found the Wigner function of the photon-added coherent state to be in agreement with that expected of an ideal cubic state, with alternating regions of positive and negative parabolic regions.

In general, the quantum state remains cubic even if the displacement is much larger, however the corresponding cubic interaction becomes weaker; we measured a fidelity of 0.995 relative to a weaker cubic state of $\gamma = 0.1$ at $\abs{\alpha} = 1.42$ (Fig. 2), with the Wigner function found to have reduced negativity as well as less oscillations across phase space. On the other hand, if the displacement is reduced, then the state deviates significantly from a cubic state and the fidelity rapidly drops. The reason for this is quite obvious: the state reduces to a Fock state in the limit of vanishing displacement.

\begin{figure}[t]
\includegraphics[width=0.5\textwidth]{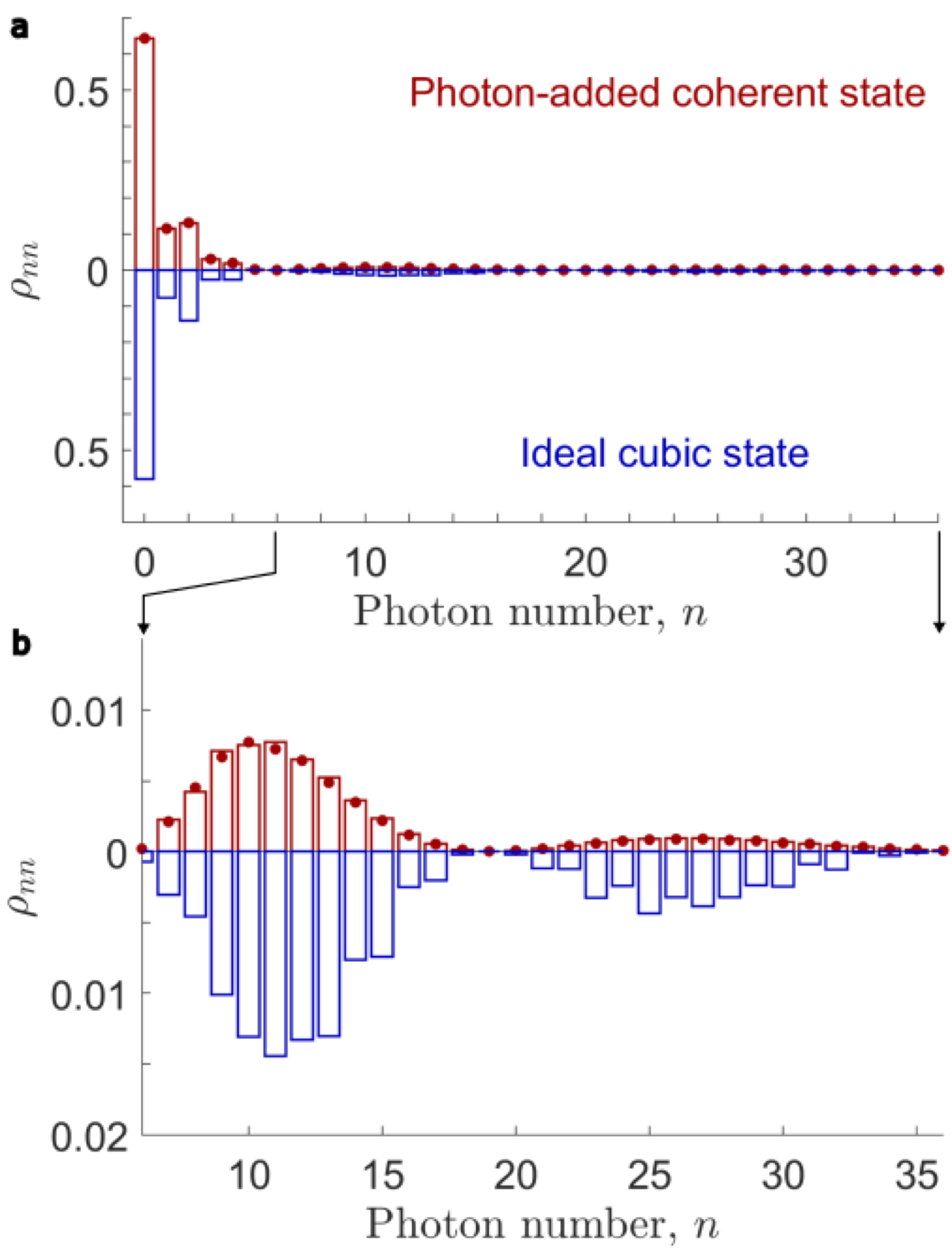}
\caption{\textbf{a.} Photon statistics of the $\gamma = 0.4$ cubic state for up to 36 photons. Red solid points represent experimental values, fitted with a model assuming only an ideal photon-added coherent state (red bars). Inverted blue bars correspond to the ideal cubic state. \textbf{b.} Magnified view of the region $6\leq n \leq 36.$}
\label{fig3}
\end{figure}

It must be emphasized that fidelity plays only a supplementary role in the identification of cubic states, with the overall similarity of Wigner functions employed as the primary method. As Yukawa \textit{et al.} have previously pointed out, the fidelity between quantum states and cubic states tend to be dominated by the largest peak in the Wigner function, neglecting other critical non-classical features such as negativities when compared to the contributions from this peak. This peak has considerable overlap with the vacuum state even for strong cubic interactions.

Examination of photon statistics reveals that the interaction strength obtained in this way is non-perturbative (Fig.~\ref{fig3}). Perturbative cubic interactions will produce states with at most three photons, Eq.~\eqref{eq3}, whereas we observed (after removing any displacement or squeezing that is present on the state by data processing) not only the existence of non-zero components beyond three photons, but also that the general structure of the photon statistics of cubic states up to at least 36 photons is reproduced. Although these probabilities for higher photons are very small, they are nonetheless visible and reveal the characteristic structure of “islands” of non-zero probabilities separated by zeroes at certain points. Analogous to how Poissonian statistics characterises coherent states and how the lack of odd photon numbers characterises squeezed states, these island structures are characteristic of the cubic phase shift.

The observations above show that cubic phase shifts produced by the method of photon addition are at least an order of magnitude stronger than those obtained by Yukawa \textit{et al.} The numerical value of the interaction strength $\gamma$ depends on the value of $\hbar$ through the definition of $\hat{x}$; we use $\hbar=2$ (corresponding to the variance of the vacuum state being equal to 1), so that the interaction strength obtained in Yukawa’s experiments becomes $\gamma = 0.035$ when converted into our units.

The ambiguity in the definition of $\gamma$ explains why it is necessary to use photon statistics to tell whether the state is non-perturbative, even though the definition for non-perturbative is a \textquote{sufficiently large $\gamma$}. For any value of $\gamma$ can be mapped to any other just by a change in the units of $\hbar$, while the photon statistics on the other hand give the probability of detecting $n$ photons from the given quantum state and therefore does not depend on the definition of $\gamma$.

\begin{figure*}[t]
\includegraphics[width=\textwidth]{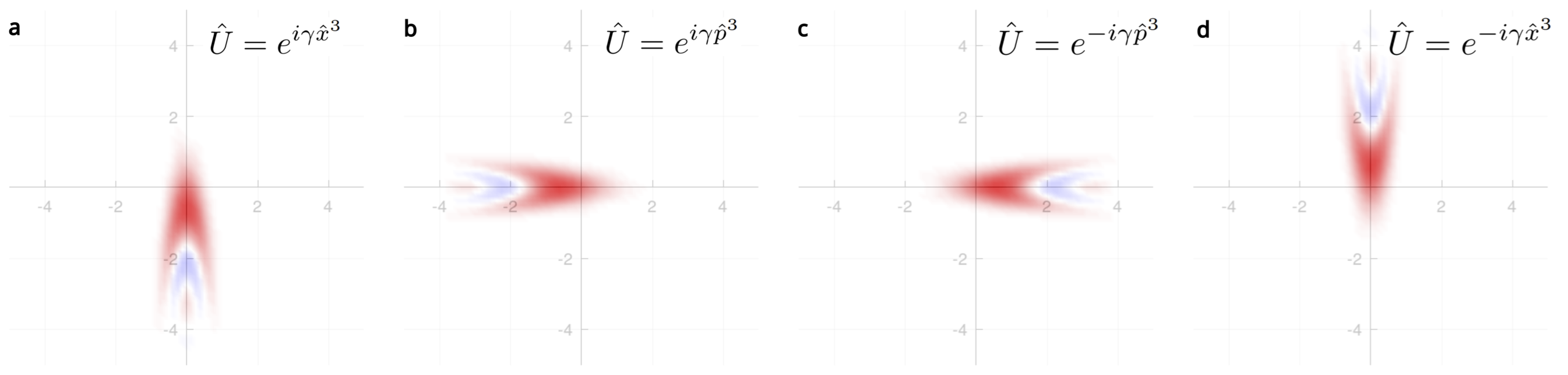}
\caption{Reconstructed Wigner functions with the initial coherent states having a fixed magnitude of approximately $\abs{\alpha}=1$ but varying phases; $\alpha = 0.01 - 0.97\mathrm{i}$, $1.06 - 0.01\mathrm{i}$, $-1.06 + 0.01\mathrm{i}$, $0.93\mathrm{i}$ respectively. The operator $\hat{U}$ indicates the phases of the corresponding cubic unitary, with an interaction strength of $\gamma=0.4$. The fidelities are $0.943$, $0.927$, $0.937$, and $0.947$ respectively.}
\label{fig4}
\end{figure*} 

As a matter of completeness, Fig. 4 shows that various phases of the cubic interaction can be obtained simply by varying the phase of the displacement. Experiments with displacements of roughly $1$, $\mathrm{i}$, $-\mathrm{i}$, and -1 were found to correspond respectively to $\text{exp}(\mathrm{i}\gamma\hat{p}^3)$, $\text{exp}(-\mathrm{i}\gamma\hat{x}^3)$, $\text{exp}(\mathrm{i}\gamma\hat{x}^3)$ and $\text{exp}(-\mathrm{i}\gamma\hat{p}^3)$ , as one would expect (with $\hat{p}$ denoting the phase quadrature). By extension, any linear combination of $\hat{x}$ and $\hat{p}$, $\hat{U}=\text{exp}(\mathrm{i}\gamma(\hat{x}\text{cos}\theta+\hat{p}\text{sin}\theta)^3)$, can be implemented by this method.

Taking the various observations together, we can summarize the relationship between photon-added coherent states and the corresponding cubic state as follows: (i) The strength of the cubic phase shift varies inversely with the magnitude of the displacement, and (ii) the phase of the cubic phase shift is in one-to-one correspondence with the phase of the displacement. These statements hold provided the displacement is not too small, as the states turn into Fock states in this final stage of transformation, first described in detail in the experiments of Zavatta, Viciani, and Bellini~\cite{ref13}.

It has not escaped our attention that the \textquote{quantum-to-classical transition}, described by Zavatta \textit{et al.}, from classical wave-like coherent states into quantum particle-like Fock states in actuality takes place in two steps, passing through an intermediate state possessing quantum, wave-like properties that closely resembles the cubic state. These observations thus complement the findings of Zavatta \textit{et al.}

In the remainder of this paper, we wish to turn our attention to the various loose ends that invariably arises during an investigation of this kind. To begin with, our method permits addition of arbitrarily many photons, and indeed, cubic phase shifts will also be generated in a similar fashion with the added photons creating ripples and distortions in the Wigner function; the number of ripples depend on the number of photons added. To simplify the investigation, we fixed the number of photons added to three because this was deemed to be sufficient for observing strong interactions while also being sufficiently practical. It is not possible to produce strong cubic phase shifts with fewer added photons (c.f. the experiments of Zavatta \textit{et al.}~\cite{ref13}), and the probability of success will be further reduced if more photons was added, by a factor of approximately 1/10 for each added photon.

To further increase the strength of the cubic interaction, it would seem natural to make use of displaced, amplitude-squeezed light in place of coherent states, owing to its elongated Wigner function which could support greater curvatures after photon addition. However, we must await the results of further investigations for a detailed comparison between states generated by such a technique and the true cubic states.

The destructive nature of the postselection method used to implement photon addition necessarily implies that the cubic states cannot be used for further quantum information processing---as ancillary states to measurement-based cubic gates, for instance. Nevertheless, this is a limitation of the specific photon addition implementation used here and not of the method itself. Provided physical sources of single photons are used, freely propagating cubic states can be generated using photon addition. We believe, however, that notwithstanding the destructive nature of our method, the observations reported here provide valuable insights for further experiments on the subject of cubic phase shifts.

It is important to note that certain schemes for multi-photon addition require photon number resolving detectors, most notably those that depend on spontaneous parametric down conversion~\cite{fadrny2024experimental}. We would like to point out that this is not a fundamental necessity, and that there are other methods of photon addition that does not require such an assumption~\cite{arend2024electrons}.

The method given here is probabilistic, and that remains true regardless of the implementation of photon addition. Although the cubic interaction is not required by quantum mechanics to be probabilistic since it is a unitary operation (a deterministic method has already been suggested theoretically by Yanagimoto \textit{et al.}~\cite{ref9}), so far as we know there does not exist any nonlinear optical interaction sufficiently strong for the realisation of deterministic cubic phase shifts.

Budinger, Furusawa, and van Loock~\cite{ref14} have investigated theoretically an alternative route of universal continuous-variable quantum computing that uses only weak cubic interactions, which at times may even be fixed and non-variable. Such recourse would be made necessary only in the situation where cubic states could not be easily produced. This fact illustrates the advantages of the photon-addition method: attainment of greater interaction strengths implies that only one interaction is needed where many sequential weak interactions may be required, and the ease with which the strength and phase could be varied indicates greater flexibility and possibilities in the design of optical quantum computation schemes.

This work was supported by the Australian Research Council (ARC) under the Centre of Excellence for Quantum Computation and Communication Technology (Grant No. CE170100012).

\bibliography{cubic_bib}
\end{document}